\documentclass[a4paper,twoside]{article}
\newcommand{\ms}{$M_{\odot}$}
\newcommand{\msb}{$M_{\odot}~$}

\newcommand{\affil}[1]{$^{\rm #1}$}
\usepackage[authoryear]{natbib}
\bibpunct{(}{)}{;}{a}{}{,}
\usepackage{graphicx}
\date{} %Please leave the date blank
\title{\large\bf\flushleft Extra-Mixing in Luminous Cool Red Giants.\\
Hints from Evolved Stars with and without Li.}
\author{\parbox{\textwidth}{\flushleft
\vspace{-0.5cm}
{\it R. Guandalini\affil{A,B,D},  S. Palmerini\affil{A,B}, M. Busso\affil{A,B}, and S. Uttenthaler\affil{C}}\\
\vspace{0.4cm}
{\small \affil{A}\,Dipartimento di Fisica, Universit$\rm\grave{a}$ degli Studi di Perugia, via Pascoli, 06123 Perugia, Italy}\\
{\small \affil{B}\,INFN Sezione di Perugia, via Pascoli, 06123 Perugia, Italy}\\
{\small \affil{C}\,Instituut voor Sterrenkunde, K. U. Leuven, Celestijnenlaan 200D, 3000 Leuven, Belgium}\\
{\small \affil{D}\,Email: guandalini@fisica.unipg.it}}}
\begin{document}
\twocolumn[
\begin{changemargin}{.8cm}{.5cm}
\begin{minipage}{.9\textwidth}
\vspace{-1cm}
\maketitle
%
%
%%%%%%%%%%%%%     ABSTRACT    %%%%%%%%%%%%%
%Abstract of no more than 200 words here.
\small{\bf Abstract: We present an analysis of Li abundances in
low mass stars (LMS) during the Red Giant Branch (RGB) and
Asymptotic Giant Branch (AGB) stages, based on a new determination
of their luminosities and evolutionary status. By applying
recently suggested models for extra-mixing, induced by magnetic
buoyancy, we show that both Li-rich and Li-poor stars can be
accounted for. The simplest scenario implies the development of
fast instabilities on the RGB, where Li is produced. When the
fields increase in strength, buoyancy slows down and Li is
destroyed. $^3$He is consumed, at variable rates. The process
continues on the AGB, where however moderate mass circulation
rates have little effect on Li due to the short time available.
O-rich and C-rich stars show different histories of Li
production/destruction, possibly indicative of different masses.
More complex transport schemes are allowed by magnetic buoyancy,
with larger effects on Li, but most normal LMS seem to show only
the range of Li variation discussed here}.

%%%%%%%%%%%%%     KEYWORDS    %%%%%%%%%%%%%
\medskip{\bf Keywords:} Stars: evolution; Stars: Red giants; Stars: AGB; Nucleosynthesis; Infrared Photometry
% Please write all keywords in lower case. PASA uses the
% standard list of subject headings adopted by The Astrophysical Journal
% and available from http://www.journals.uchicago.edu/ApJ/keywords_text.html.
% Keywords are separated by em-dashes, i.e. ---

%%%%%%%%DO NOT EDIT%%%%%%%%%%%%
\medskip
\medskip
\end{minipage}
\end{changemargin}
]
\small
%%%%%%%%EDIT FROM HERE%%%%%%%%%%%%

\section{Introduction}
Most K-type giants are depleted in lithium. For low-metallicity
objects an upper limit on the Li content was set by
\citet{gratton}, stating that field, metal-poor (hence low-mass)
stars on the upper RGB have $\log~ \epsilon$(Li) $\le$ 0

Mixing processes on the Main Sequence and up to the first
dredge-up imply that Li be gradually carried down from the
photosphere to regions of high temperature and destroyed, while
any amount of $^7$Be produced by H burning burns "on the flight"
and new Li is not produced. The observations of Li-poor red giants
are therefore well understood. It was further established three
decades ago \citep{sweigart} and supported by many observations
\citep[e.g.][]{gilroy,gilroybrown,kraft} that, after the first
dredge-up, low mass stars must experience other mixing episodes,
sufficiently slow to gradually carry to the surface a considerable
amount of $^{13}$C, so that the $^{12}$C/$^{13}$C ratio decreases
from 25-30 at the first dredge-up, down to 12-15 (in population I
stars) or to 4-8 (in population II red giants). The occurrence of
these mixing phenomena is made easy by the fact that, after
dredge-up, the H burning shell advances in a homogeneous region,
so that the natural barrier opposed to mass transport by a
chemical stratification is not present. The onset of this phase is
accompanied by a local decrease in the luminosity, followed by a
new rise, so that the same regions in the H-R diagram are crossed
repeatedly and the points representing coeval stars pile up,
producing a bump in the Luminosity Function. This is the so-called
"L-bump phase".

Any slow mixing process occurring at or after the L-bump would
further reduce not only Li, but also $^9$Be and $^3$He
\citep{bs99}, as confirmed by observations \citep{cast}. However,
a few red giants (about 2\%) show enhanced Li abundances, at
levels sometimes higher than in the present Interstellar Medium
($\log~\epsilon$(Li) $\simeq 3.3$). Various explanations have been
attempted, the most common assumption being that some form of fast
extra-mixing might transport $^{7}$Be from above the H-burning
shell to the envelope at a speed sufficient to overcome the rate
of p-captures; this is the so-called Cameron-Fowler mechanism
\citep{camfow}. Further production of $^7$Be along the path can
also occur \citep{bs99}. Very often, the cause of such
extra-mixing was looked after in the stellar rotation, inducing
shear instabilities, meridional circulation and diffusion
phenomena \citep{charb94,charb98,denvan,palacios03}. These
possibilities were frustrated by the understanding that the
stellar structure reacts to rotational distortions too quickly to
induce significant mixing on long time scales \citep{palacios06}.

Recently, two old ideas on mass transport have been revisited,
looking for alternative causes of deep circulation in stars. Both
attribute mass transport to the presence, in deep layers, of
materials lighter than in the envelope. The first mechanism ({\it
thermohaline} diffusion) attributes this to the activation of the
reaction $^3$He+$^3$He $->$ $^4$He + 2p \citep{eggleton}, reducing
the molecular weight. The second process is the buoyancy of
magnetized H-burning ashes \citep{busso07a,wasserburg,nordhaus},
based on the fact that magnetic bubbles are lighter than the
environment, due to the unbalance generated by magnetic pressure
($B^2/8\pi$). This requires a magnetic dynamo to occur in red
giants, as indeed observed \citep{and88}.

Triggering these mechanisms requires specific conditions: the
first one needs a suitable molecular weight inversion; the second
is possible if strong magnetic fields exist. The two processes
might be complementary and even occur together \citep{den09}.
However, while thermohaline mixing implies a diffusive, slow
transport, magnetic buoyancy can occur at different speeds,
depending on the amount of heat exchange with the environment
\citep{den09}. Therefore, a possibility of understanding which
mechanism is at play might be provided by observations of nuclei,
like Li, for which the production and destruction are sensitive to
the mixing velocity.

Such a test requires an accurate calibration of the luminosity of
Li-bearing stars, in order to attribute them to the proper
evolutionary stage. Therefore, we here apply bolometric
corrections for evolved stars, as derived in our recent work, to
infer the absolute HR diagram of red giants showing Li in their
spectrum. This helps in setting constraints on where extra-mixing
is active. In Section 2 we present a sample of cool red giants
showing highly dispersed Li abundances, ranging from very Li-rich
to very Li-poor, and we determine for them absolute magnitudes
with the help of our bolometric corrections and of recent distance
estimates. In Section 3 we compare the inferred magnitudes with
the Li content, as taken from the literature and we present
results of nucleosynthesis and mixing calculations, using the
transport velocities suggested by magnetic buoyancy. Finally, in
Section 4 preliminary conclusions are illustrated.

\section{Bolometric Corrections and Li-abundances in Red Giants}

Any growth in our understanding of AGB stars is today bound to
improvements on two crucial parameters: luminosities and mass loss
rates. For both, determinations are hampered by the large distance
uncertainties; both are also affected by the fact that most of the
radiation emitted by these sources can be observed only at
infrared wavelengths and therefore depends on space-borne
facilities, or maybe on the exploitation of special ground-based
locations, like Antarctica \citep{irait}.

In the last years we performed an extensive analysis of the
photometric properties of AGB stars, using available infrared data
from the ISO and MSX experiments and looking for relations between
their luminosities and their main chemical and physical
parameters. The aim of this work was to put constraints on basic
stellar parameters, thus offering calibrating tools to
evolutionary models. In this framework, large samples of C-rich,
S-type, M-type AGB stars were collected and studied. The first
results were published for carbon-rich and S-type stars in
\citet{guabus,busso07b,gua06}. In Figures 1 and 2 we present the
bolometric corrections we derived for samples of galactic C-rich
and O-rich (S-type) stars, respectively. The techniques adopted
and the source properties are described in \citet{gua06} and in
\citet{guabus}.

For the purposes of the present work we use the bolometric
corrections of Figures 1 and 2, applying them to a sample of AGB
stars showing highly dispersed Li abundances (from Li-rich to
Li-poor). For these sources we determine the apparent bolometric
magnitude and we use the recent revision of the Hipparcos
catalogue \citep{hipparcos}, to infer the absolute magnitude. The
AGB stars considered, their magnitudes and their Li abundances are
shown in Table 1. In Table 2 we present a second sample, made of
RGB stars. The bolometric magnitudes are taken from the
literature, with distances updated by \cite{hipparcos}. Some
explanations can be useful in presenting the sources in Tables 1
and 2.

\begin{figure}[t*]
\begin{center}
\includegraphics[scale=0.4, angle=0]{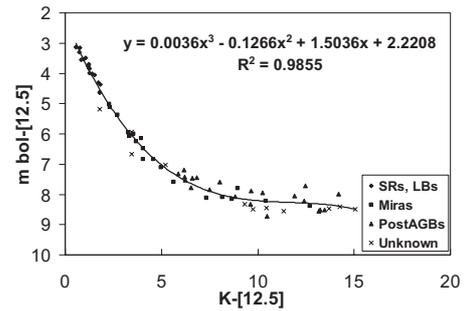}
\caption{The bolometric correction for C stars, as a function of the
infrared color K-[12.5].}\label{fig1}
\end{center}
\end{figure}

\begin{figure}[t*]
\begin{center}
\includegraphics[scale=0.4, angle=0]{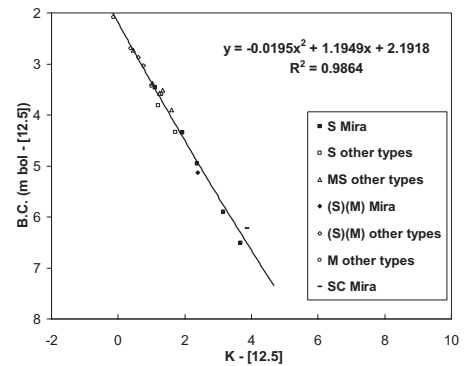}
\caption{The bolometric correction for O-rich stars, derived from
MS- and S-type red giants, as a function of the infrared color
K-[12.5].}\label{fig2}
\end{center}
\end{figure}
A) Horizontal lines in the tables divide the stars of the sample
in a few sub-groups, according to their spectral classification;
from top to bottom we find:

\onecolumn
%%Format tables as in the following example
\begin{table}[t*]
\begin{center}
\caption{AGB sample}\label{table1}
\begin{tabular}{llcrrcl}
\hline Source Name & Sp. Type & D (kpc)  & m$_{bol}$ & M$_{bol}$ & $\delta$M$_{bol}$ & Log $\varepsilon$(Li)  \\
\hline\hline
WZ Cas  &   C9,2JLi(N1P)    &   0.88    &   3.16    &   $-$6.56   &   1.26    &   +5.80$^a$$^,$$^b$ \\
Y CVn   &   C5,4J(N3)   &   0.32    &   1.73    &   $-$5.80   &   0.49    &   $-$0.40$^a$$^,$$^b$    \\
RY Dra  &   C4,5J(N4P)  &   0.43    &   3.10    &   $-$5.07   &   0.80    &   $-$0.45$^a$$^,$$^b$   \\
T Lyr   &   C$-$J4:p C2 5 j3.5    &   0.72    &   2.86    &   $-$6.43   &   1.02    &   +2.20$^a$ \\
VX And  &   C$-$J4.5 C2 5 j5 MS5  &   0.39    &   3.88    &   $-$4.09   &   0.93    &   +2.60$^c$ \\
V614 Mon    &   C4,5J(R5)   &   0.47    &   4.43    &   $-$3.95   &   1.23    &   +1.30$^c$ \\
\hline
Z Psc   &   C7,2(N0)    &   0.38    &   3.53    &   $-$4.40   &   0.78    &   $-$1.30$^a$$^,$$^b$    \\
R Scl   &   C6,5EA(NP)  &   0.27    &   2.55    &   $-$4.61   &   0.68    &   $-$0.40$^a$$^,$$^b$    \\
U Cam   &   C3,9$-$C6,4E(N5)  &   0.97    &   3.25    &   $-$6.69   &   1.50    &   $-$0.75$^a$$^,$$^b$   \\
ST Cam  &   C5,4(N5)    &   0.99    &   3.21    &   $-$6.76   &   1.65    &   $-$0.50$^a$$^,$$^b$    \\
R Lep   &   C7,6E(N6E)  &   0.41    &   2.88    &   $-$5.20   &   1.17    &   $-$1.10$^a$$^,$$^b$    \\
W Ori   &   C5,4(N5)    &   0.38    &   2.19    &   $-$5.70   &   1.03    &   $-$1.05$^a$$^,$$^b$   \\
Y Tau   &   C6.5,4E(N3) &   0.36    &   3.07    &   $-$4.72   &   1.03    &   $-$0.80$^a$$^,$$^b$    \\
TU Gem  &   C6,4(N3)    &   0.52    &   3.55    &   $-$5.03   &   0.68    &   $-$0.55$^a$$^,$$^b$   \\
BL Ori  &   C6,3(NB,TC) &   0.60    &   3.37    &   $-$5.52   &   0.68    &   $-$1.10$^a$$^,$$^b$    \\
W CMa   &   C6,3(N) &   0.79    &   3.54    &   $-$5.95   &   0.68    &   $-$1.20$^a$$^,$$^b$    \\
X Cnc   &   C5,4(N3)    &   0.34    &   2.71    &   $-$4.94   &   0.83    &   $-$1.00$^a$$^,$$^b$  \\
Y Hya   &   C5,4(N3P)   &   0.39    &   3.19    &   $-$4.77   &   0.63    &   $-$0.90$^a$$^,$$^b$    \\
VY Uma  &   C6,3(N0)    &   0.38    &   3.15    &   $-$4.75   &   0.54    &   $-$0.50$^a$    \\
S Sct   &   C6,4(N3)    &   0.39    &   3.30    &   $-$4.66   &   0.73    &   $-$0.50$^a$$^,$$^b$    \\
V Aql   &   C5,4$-$C6,4(N6)   &   0.36    &   2.55    &   $-$5.23   &   0.79    &   $-$0.65$^a$$^,$$^b$   \\
UX Dra  &   C7,3(N0)    &   0.39    &   3.03    &   $-$4.90   &   0.49    &   +0.10$^a$$^,$$^b$     \\
AQ Sgr  &   C7,4(N3)    &   0.33    &   3.51    &   $-$4.11   &   0.74    &   $-$0.95$^a$$^,$$^b$   \\
V460 Cyg    &   C6,4(N1)    &   0.62    &   2.89    &   $-$6.08   &   0.77    &   $-$0.90$^a$$^,$$^b$    \\
TX Psc  &   C7,2(N0)(TC)    &   0.28    &   2.02    &   $-$5.21   &   0.48    &   $-$0.65$^a$$^,$$^b$   \\
CR Gem  &   C8,3E(N)    &   0.35    &   4.28 &   $-$3.44   &   1.78    &   +0.00$^a$   \\
RS Cyg  &   C8,2E(N0PE) &   0.47    &   3.99 &   $-$4.39   &   1.17    &   $-$0.50$^a$    \\
RV Mon  &   C4,4$-$C6,2(NB/R9)    &   0.52    &   4.56  &   $-$4.02   &   0.99    &   $-$1.00$^a$  \\
RY Mon  &   C5,5$-$C7,4:(N5/R)    &   0.43    &   3.66 &   $-$4.52   &   0.97    &   +0.50$^a$     \\
S Cep   &   C7,4E(N8E)  &   0.41    &   2.70  &   $-$5.35   &   0.82    &   +0.00$^a$   \\
TT Cyg  &   C5,4E(N3E)  &   0.56    &   4.57  &   $-$4.18   &   0.87    &   +0.70$^a$     \\
TU Tau  &   C$-$N4.5 C2 6 &   0.43    &   4.30 &   $-$3.89   &   1.35    &   $-$0.50$^a$    \\
V Cyg   &   C5,3E$-$C7,4E(NPE)    &   0.37    &   2.74 &   $-$5.08   &   1.51    &   $-$1.00$^a$  \\
\hline
Y Lyn   &   M6SIB$-$II    &   0.25    &   1.66    &   $-$5.33   &   0.77    &   $-$2.00$^d$  \\
RS Cnc  &   M6EIB$-$II(S) &   0.14    &   0.52    &   $-$5.21   &   0.41    &   $-$2.00$^d$  \\
V1981 Cyg   &   S4/1III &   0.30    &   3.43    &   $-$3.96   &   0.46    &   $-$2.00$^d$  \\
\hline
bet And &   M0+ IIIa    &   0.06    &   0.17    &   $-$3.74   &   0.32    &   $-$0.48$^e$   \\
CU Dra  &   M3III   &   0.11    &   2.09    &   $-$3.19   &   0.30    &   $-$1.50$^e$    \\
BY Boo  &   M4$-$4.5III   &   0.16    &   2.15    &   $-$3.84   &   0.34    &   $-$0.87$^e$   \\
OP Her  &   M5IIB$-$IIIA(S)   &   0.30    &   2.47    &   $-$4.92   &   0.42    &   $-$2.00$^d$  \\
\hline
V2652 Sgr   &  M9     &   $\simeq$ 8.0   &   9.98    &   $-$4.52   &  0.30     &   +2.00$^f$   \\
V3252 Sgr   &  M7     &  $\simeq$ 8.0   &    9.74   &   $-$4.76   &   0.30    &   +1.10$^f$ \\
V3537 Sgr   &  M9S     &  $\simeq$ 8.0   &   9.22    &   $-$5.28   &  0.30     &   +0.80$^f$ \\
V2017 Sgr   &  M7S     &  $\simeq$ 8.0   &   9.07    &   $-$5.43   &  0.30     &   +0.80$^f$ \\
\hline\hline
\end{tabular}
\medskip\\
For Log $\varepsilon$(Li): $^a$\citet{boffin}, $^b$\citet{denn},
$^c$\citet{abia}, $^d$\citet{vanture},
$^e$\citet{luck},$^f$\citet{utt},
$^g$\citet{lambert},$^h$\citet{mallik},
$^i$\citet{charb00}, $^j$\citet{brown}.\\
\end{center}
\end{table}

\begin{table}[t*]
\begin{center}
\caption{RGB sample}\label{table2}
\begin{tabular}{llcrrll}
\hline Source Name & Sp. Type & D (kpc)  & m$_{bol}$ & M$_{bol}$ & T$_{eff}$ (K) & Log $\varepsilon$(Li)  \\
\hline\hline
alpha Boo   &   K1.5III &   0.01    &   $-$0.82   &   $-$1.08   &   4290    &   $-$1.50$^g$    \\
beta UMi    &   K4III   &   0.04    &   1.02    &   $-$2.00   &   4000 &   $-$1.50$^g$    \\
alpha Ser   &   K2IIIb  &   0.02    &   1.92    &   0.14    &   4300 &   $-$0.28$^g$   \\
gamma Dra   &   K5III   &   0.05    &   1.18    &   $-$2.20   &   4000 &   $-$0.80$^g$$^,$$^h$    \\
\hline
HD 787  &   K4III   &   0.19    &   4.15 &   $-$2.25   &   3990$^1$    &   +1.80$^i$ \\
HD 30834    &   K3III   &   0.18    &   3.81 &   $-$2.52   &   3920$^1$    &   +1.80$^i$ \\
HD 39853    &   K5III   &   0.20    &   4.53 &   $-$2.01   &   3920$^1$    &   +2.80$^i$ \\
HD 121710   &   K3III   &   0.18    &   4.05 &   $-$2.20   &   4100$^2$    &   +1.50$^i$ \\
HD 146850   &   K3IIICNpv   &   0.33    &   4.98 &   $-$2.59   &   4200$^3$    &   +2.00$^i$   \\
HD 9746 &   K1III:  &   0.16    &   5.51 &   $-$0.47   &   4420$^2$    &   +3.75$^i$    \\
HD 148293   &   K2III   &   0.09    &   4.71 &   $-$0.06   &   4640$^2$     &   +2.00$^i$   \\
HD 183492   &   K0III   &   0.09    &   5.10  &   0.29    &   4720$^1$    &   +2.00$^i$   \\
HD 219025   &   K2IIIp  &   0.26    &   7.03 &   $-$0.06   &       &   +3.00$^i$   \\
\hline
HD 3817 &   G8III   &   0.11    &   4.85 &   $-$0.34   &   5020$^2$    &   $<$+0.70$^j$ \\
HD 3856 &   G9III$-$IV    &   0.16    &   4.90 &   $-$1.10   &   4750$^2$    &   +1.20$^j$ \\
HD 4627 &   G8III   &   0.18    &   5.28   &   $-$0.96   &   4610$^2$    &   $<$+0.00$^j$   \\
HD 19845    &   G9III   &   0.12    &   5.33 &   0.00    &   4830$^2$    &   $<$+0.30$^j$ \\
HD 78235    &   G8III   &   0.08    &   5.08 &   0.46    &   5000$^2$    &   +0.80$^j$ \\
HD 108225   &   G9III   &   0.08    &   4.66 &   0.14    &   4960$^2$    &   $<$+0.60$^j$ \\
HD 119126   &   G9III   &   0.10    &   5.14 &   0.07    &   4750$^2$    &   $<$+0.20$^j$ \\
HD 120048   &   G9III   &   0.13    &   5.49 &   $-$0.09   &   4890$^2$    &   +1.00$^j$  \\
HD 129336   &   G8III   &   0.12    &   5.09 &   $-$0.37   &   4940$^2$    &   $<$+0.50$^j$ \\
HD 134190   &   G7.5III &   0.08    &   4.75 &   0.26    &   4880$^2$    &   $<$+0.40$^j$ \\
HD 141680   &   G8III   &   0.08    &   4.59 &   $-$0.02   &   4750$^2$    &   $<$+0.20$^j$ \\
HD 152815   &   G8III   &   0.08    &   4.90 &   0.31    &   4900$^2$    &   $<$+0.40$^j$ \\
HD 158974   &   G8III   &   0.12    &   5.14 &   $-$0.28   &   4890$^2$    &   $<$+0.40$^j$ \\
HD 163532   &   G9III   &   0.12    &   4.40 &   $-$1.04   &   4710$^2$    &   $<$+0.20$^j$ \\
HD 171391   &   G8III   &   0.10    &   4.44 &   $-$0.58   &   4980$^2$    &   +1.20$^j$ \\
HD 176598   &   G8III   &   0.10    &   5.25 &   0.34    &   4925$^2$    &   +1.10$^j$ \\
HD 186675   &   G7III   &   0.09    &   4.49 &   $-$0.24   &   4910$^2$    &   $<$+0.30$^j$ \\
HD 192944   &   G8III   &   0.13    &   4.77 &   $-$0.87   &   4900$^2$    &   $<$+0.30$^j$ \\
HD 194013   &   G8III$-$IV    &   0.08    &   4.86 &   0.46    &   4850$^2$    &   $<$+0.40$^j$ \\
HD 201381   &   G8III   &   0.05    &   4.04 &   0.60    &   4960$^2$    &   $<$+0.50$^j$ \\
HD 211391   &   G8III   &   0.06    &   3.68 &   $-$0.11   &   4900$^2$    &   $<$+0.30$^j$ \\
HD 211554   &   G8III   &   0.19    &   4.87 &   $-$1.51   &   4920$^2$    &   +0.50$^j$ \\
HD 215030   &   G9III   &   0.09    &   5.41 &   0.60    &   4800$^2$    &   +0.60$^j$ \\
\hline\hline
\end{tabular}
\medskip\\
For T$_{eff}$: $^1$\citet{melo}, $^2$\citet{brown},
$^3$\citet{charb00}.\\ For Log $\varepsilon$(Li):
$^a$\citet{boffin}, $^b$\citet{denn}, $^c$\citet{abia},
$^d$\citet{vanture}, $^e$\citet{luck},$^f$\citet{utt},
$^g$\citet{lambert},$^h$\citet{mallik},
$^i$\citet{charb00}, $^j$\citet{brown}.\\
\end{center}
\end{table}

\twocolumn

\begin{itemize}
    \item In Table 1: 1) CJ stars, 2) the C(N) sample, 3) S-type sources, 4) M-type
giants and finally 5) a few O-rich AGB sources of the Galactic
bulge from \citet{utt}.
    \item In Table 2: 1) A sample of close-by, Li-poor K-type stars, 2) a similar
    one for Li-rich K-type sources, 3) a group of G-type, Li-depleted giants.
\end{itemize}

B) As mentioned, distances (in kpc) are in general from
\cite{hipparcos}. For three sources in Table 1, for which the
Hipparcos' parallaxes are not available, we use the data from
\citet{bergeat}. For the four stars of the bulge the estimates of
the distance and therefore also the absolute bolometric magnitudes
are taken from \citet{utt}.

C) The apparent bolometric magnitude of the sour\-ces in Table 1 was
obtained through the integration of the data from ISO, 2MASS, MSX
and IRAS-LRS at different wavelengths and by applying the bolometric
corrections suitable for their chemical type. Concerning
K-type stars, due to their known distance and
luminosity, both their apparent and absolute bolometric magnitude
are taken from the available literature.

D) In Table 1 a typical uncertainty of 0.25 mag can be attributed to
apparent magnitudes; the uncertainty on the distance is given by
\citet{hipparcos}.

E) For the two sub-groups at the bottom of Table 2 (second and third
one) we calculated the magnitudes from data available in the literature
\citep{charb00,takeda}, updated with new estimates of the distance.

F) The Li abundances were taken from various references: the full lists
are shown in the footnotes to the tables. In the case of
sources with multiple references we used an average of the data available.

G) The effective temperatures presented in Table 2 come
from the list of references indicated in the footnote to the same Table.

{\begin{figure}[t!]
   \centering
   {\includegraphics[width=\columnwidth]{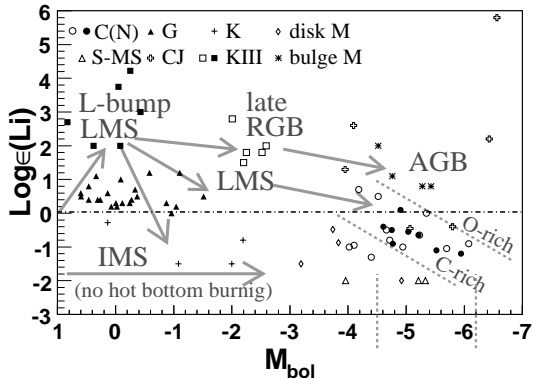}}
\caption{The sample of cool red giants showing Li in their spectra.
Arrows indicate the suggested evolutionary path induced by
extramixing. Dashed lines roughly limit the region occupied by C(N)
giants and their range of luminosities as inferred from the
empirical Luminosity Function of Figure 4. The dash-dotted line
indicates the maximum Li abundance observed in pop. II red
giants.}\label{fig3}%
    \end{figure}
}

\section{On the Luminosities of Red Giants Showing Li}

Figure 3 presents the selected sample of evolved stars showing Li
in their spectra, after correcting their luminosities with the
criteria described in the previous section. The plot shows several
groups of stars. As compared to the ISM Li abundances, many of the
stars in our sample are strongly depleted in Li
($\log~\varepsilon$(Li) $\le 1$). Two main outlying groups exist
(including mainly K giants) with Li abundances near
$\log~\varepsilon$(Li) $\simeq 2$. A few super-Li rich stars are
also present, with $\log~\varepsilon$(Li) $\ge 3$. We shall not
discuss  super-Li rich stars in this paper, as they are often
peculiar objects (like CJ giants). Their high Li abundance can be
produced either by a relatively fast extra-mixing in LMS or by Hot
Bottom Burning in more massive AGB stars.

The group of stars at the left of the plot, with
$\log~\varepsilon$(Li) $\simeq 2$, is made of K giants with
luminosities typical of the L-bump on the RGB. In agreement with
\citet{charb00} we consider them as being stars newly enriched in
Li. After the extensive depletion of Li in the Main Sequence and
in the early RGB phases, their envelopes must have seen some form
of rapid mixing implementing a Cameron-Fowler mechanism, thus
bringing to the surface fresh $^7$Be synthesized above the
H-burning shell.

The second moderately Li-rich group was interpreted by
\citet{charb00} as being formed by early-AGB stars, producing Li
(after some destruction in the late RGB phases). Inspection of
stellar models, however, reveals that the early-AGB stages
suitable to reproduce the luminosities and temperatures of these K
giants fall in a temporal phase where the H shell is extinguished,
so that no $^7$Be survives to be carried to the surface. We must
therefore tentatively conclude that a new production of Li on the
early-AGB at temperatures and luminosities compatible with
observations does not occur and therefore the observed Li must be
a relic of the previous production on the RGB. We actually
interpret this group of stars as being in the upper part of the
RGB itself (see next section).

On the right side of the plots we find AGB stars, either from the
galactic disk or from the galactic bulge: these last are O-rich,
while for the galactic disk most (although not all) Li
observations concern C-rich stars. As an eye-guide we have marked
by dashed lines the region occupied by C(N) giants, and in Figure
4 we also report their Luminosity Function, as determined by the
works mentioned in Section 2. Vertical dashed lines on the two
figures allow one to realize that Li-bearing C(N) stars belong to
the typical luminosity intervals of normal C stars. We believe
that also their Li abundances should represent typical trends.

\section{Modeling the Li Production and Destruction}

As discussed elsewhere \citep{palm2} mixing mechanisms induced by
magnetic buoyancy offer a scenario in which both fast and slow
matter circulation can occur. A simple example of fast transport
is by magnetic instabilities, when portions of a magnetic
$\Omega-$shaped loop detach and travel the radiative layers above
the H shell and the convective envelope at a velocity close to the
Alfv\'en speed. In this case Be is produced by normal H-burning
above the H-shell, refueled by fresh $^3$He carried down by the
mixing. When fields grow above a certain limit, buoyant
instabilities are released sporadically, transporting (without
local mixing) their composition to the surface. This form of Li
production has an intrinsic limit in the fact that no burning
occurs along the path, so that the maximum Li abundance reachable
is set by the equilibrium Be concentration near the H shell (this
corresponds to about $\log \varepsilon_{Max}(Li) = 2 - 2.5$).

Alternatively Li can be produced by a slower mixing process, where
$^3$He has time to burn along the path; the mechanism must however
be still fast enough to save some of the produced Be. In general,
producing or destroying Li depends on a delicate equilibrium of at
least four different parameters: i) the velocity of mixing; ii)
its maximum penetration (hence the maximum temperature reached,
$T_P$); iii) the initial Li content; and iv) the time available
for the process.

\begin{figure}[h*]
\begin{center}
\includegraphics[scale=0.45, angle=0]{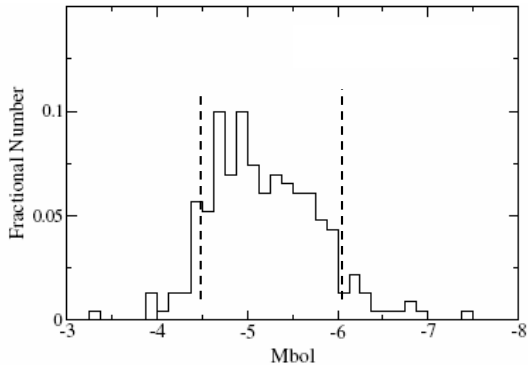}
\caption{The Luminosity Function of C(N) stars (see text).
Vertical dashed lines establish a rough fiducial interval for the
luminosities, as also indicated in Figure 3.}\label{fig4}

\end{center}
\end{figure}

We adopt the formalism by \citet{palm1}, derived from the idea
presented by \citet{busso07a} for magnetically-induced mixing.
With these tools we try here to account for the observed Li
abundances in evolved LMS.

\begin{figure}[t*!!]
\begin{center}
\includegraphics[angle=0,width=7cm]{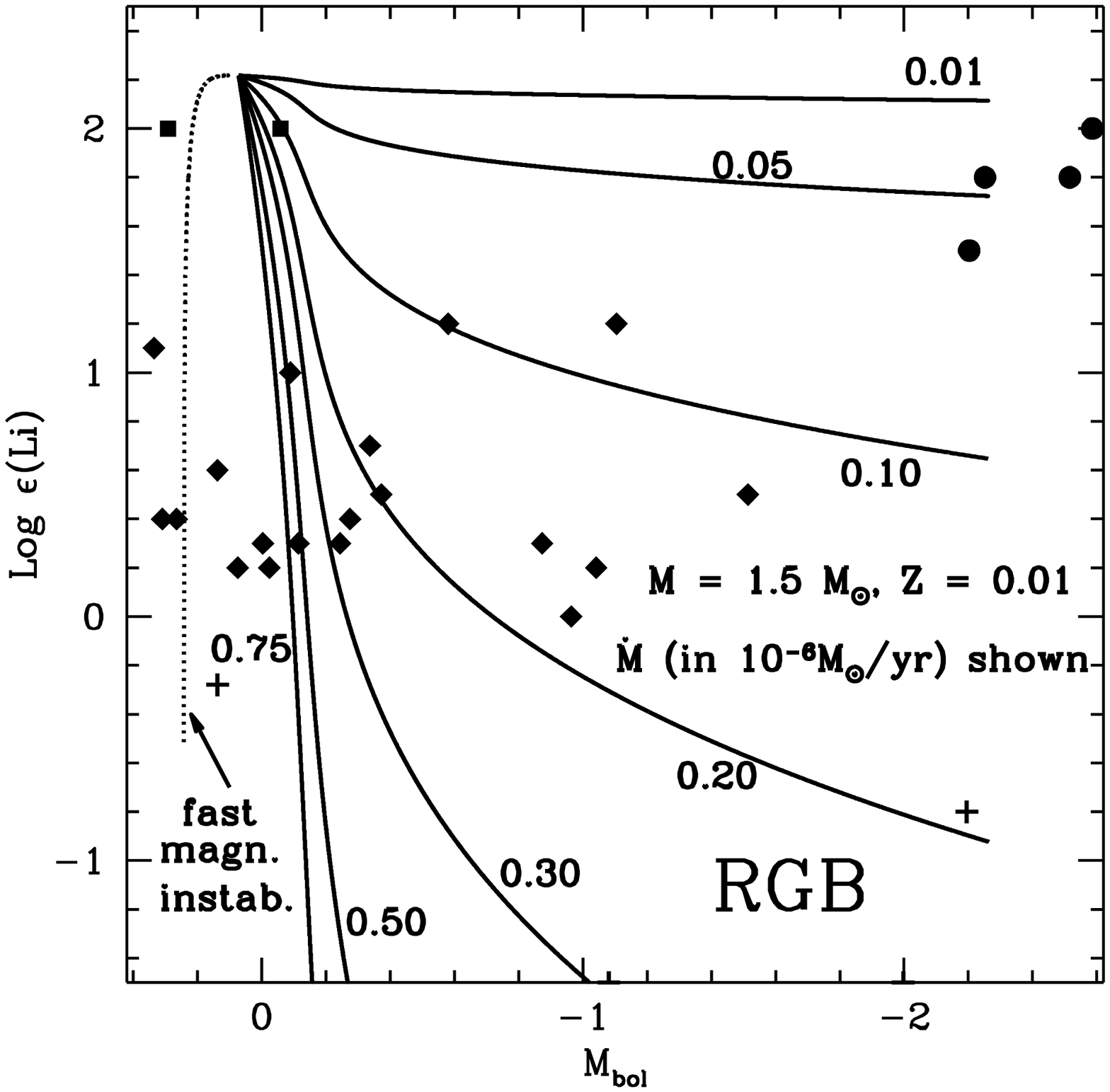}
\caption{RGB stars showing Li and model results from the mixing
processes discussed in the text. Rhombs are GIII stars from
\citet{brown,takeda}. Filled squares are stars at the L-bump from
\citet{charb00}); crosses are K giants; filled squares are RGB
stars that remained Li-rich. The dotted line indicates Li
enrichment through fast magnetic instabilities. Solid lines show
the effects of slower transport at different mixing rates. This
might be induced by the buoyancy of large structures, exchanging
heat with the environment.}\label{fig5}
\end{center}
\end{figure}

\begin{figure}[t*!!]
\begin{center}
\includegraphics[angle=0,width=7cm]{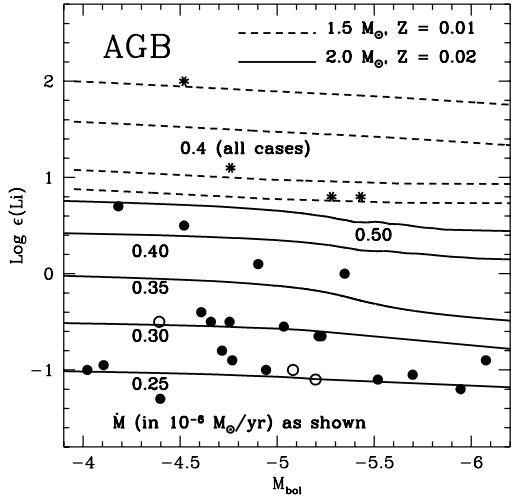}
\caption{AGB stars showing Li and model results from magnetic
mixing. Circles are C(N) stars (the empty ones with larger
luminosity uncertainties); asterisks show O-rich AGB objects in the
galactic bulge \citep{utt}: the most luminous of them also show
Tc in their spectra.}\label{fig6}
\end{center}
\end{figure}

We consider two different types of magnetic buoyancy, which can be
described as follows.

i) (\textbf{Model A}). Integrating the assumptions by
\citet{busso07a}, we consider a phase of magnetic field growth, at
the L-bump, where fields are assumed to be not strong enough to
promote the buoyancy of entire flux tubes; only local
instabilities can detach from regions near the H shell, traveling
at the Alfv\'en speed $B/\sqrt(4 \pi \rho)$, up to a few Km/sec.
In our exercise, the release of unstable magnetized bubbles is
fine-tuned to produce a rate of mass addition to the envelope of
10$^{-6}$ \ms/yr and the Alfv\'en velocity is computed adopting
field strengths from \citet{busso07a}. An equally fast downflow of
envelope material guarantees mass conservation, replenishing the
inventory of $^3$He in the radiative regions. No burning occurs in
the circulation, because the motion is very fast. The net effect,
for the production of Li, is the mixing of the "normal" $^7$Be
(produced near the H shell) into the envelope, where it will decay
to Li. Contemporarily, the $^3$He abundance at the surface
decreases.

ii) (\textbf{Model B}). As an opposite case, we consider that,
when the fields have grown sufficiently, larger structures, or
even entire flux tubes become buoyant, driving a circulation of
mass that slows down due to the gradual heat exchange between the
magnetized zones and the environment (heat exchanges grow with the
surface of the emerging structure). The speed is reduced down to
values of few cm/sec \citep{den09}. Again, $^3$He is destroyed at
various efficiencies, depending on the parameters mentioned above.
The new property of our model is that, if magnetic fields are at
the origin of the transport, then one has a justification for both
fast and slow regimes of circulation, and for both production or
destruction of Li; a single physical mechanism can explain very
different outcomes, depending on the parameters (and observations
will then provide the key information on the parameter values).
Models of extra-mixing so far presented lack this characteristic,
as diffusive processes induced by either rotational effects or
thermohaline mixing have a very low velocity.

Magnetic fields can even generate more complex schemes of mixing,
e.g. with upward and downward fluxes occupying different mass
fractions and traveling at different speeds. We shall afford a
more complete analysis, including those possibilities, in a
forthcoming paper. For the sake of illustration it can be now
sufficient to limit the discussion to the simple and rather
extreme cases of Models A and B.

As mentioned, we apply Model A at the luminosity bump, then we
switch to the slow circulation of Model B. This way of proceeding
is rather ad-hoc, although it might be qualitatively reasonable.
We do not know enough of magnetic fields in red giants to claim
that our hypotheses are really the correct ones, but they have at
least a basis in the observations, showing Li-rich stars at the
L-bump and (mostly) Li-poorer stars in more evolved stages (again
excluding super Li-rich objects). Ours is therefore an exercise,
aiming at showing how, with suitable assumptions, magnetic
buoyancy can in fact offer a framework where the puzzle of Li
production and destruction can find a solution.

With the above cautions in mind, Figures 5 and 6 present the
results of our effort. In Figure 5 we show, by a dotted line, the
effects of applying Model A over a composition typical of the
post-main sequence phase. Li is rather rapidly produced (in a few
million years) up to levels of $\log~\varepsilon$(Li) = 2 - 2.5
(depending on the depth of mixing and on the temperature
distribution). The example shown is for a 1.5 \msb model of half
solar metallicity but is rather typical of the intervals 0.01
$\le$ $Z$ $\le$ 0.02, 1.5 $\le$ $M$/\msb $\le$ 2. The solid lines
represent instead results of the application of Model B with
various (moderate) rates for mass circulation. As the Figure 5
shows, by varying this rate one can explain the whole range of Li
destruction displayed by observed red giants, and even an almost
complete Li survival, thus accounting for the few cool Li-rich
stars. This substantiates our suggestion that they are late RGB
objects where Li was not destroyed, not early-AGB stars where Li
was newly produced. The maximum temperature achieved by mixing
($T_P$) is chosen at the position where the equilibrium $^7$Be has
its maximum, near $\log T_{\rm H} - \log T_P$ = 0.3. One can
notice that, for this $T_P$ value, the destruction of Li becomes efficient
only for the higher $\dot M$ values shown. However, if $\dot M$
were further increased one might obtain an excessively fast
circulation, and Li would be produced. Li production occurs when
$\dot M$ is above a limiting value, somewhere between 10$^{-6}$
and 10$^{-5}$ M$_{\odot}$/yr, the actual value depending on $T_P$
and on the time available for the mixing to operate. The large
number of different possible outcomes thus forms a real zoo, in
which only observations can effectively constrain the parameters.

Finally, Figure 6 illustrates the effects of applying Model B,
with moderate mass circulation rates, to AGB stars of 1.5 - 2.0
\ms. No change would be seen by adopting the same rule for $T_P$
used for the RGB ($\log T_{\rm H} - \log T_P$ = 0.3) (the time
available is too short on the AGB). We have therefore pushed $T_P$
to its maximum possible value that still does not disturb the
stellar equilibrium ($\log T_{\rm H} - \log T_P$ = 0.1). Even so,
with moderate transport rates the changes are rather small and
would indicate that most of the Li production or destruction has
occurred previously, on the RGB. Stronger transport rates would
induce Li production also on the AGB, but they might also imply
strong difficulties in saving enough carbon to allow the stars to
become C-rich. These are general indications, not firm
conclusions, as we have seen that several parameters affect Li
abundances.

In conclusion our results can be summarized as follows:

i) Outlining the details of Li production and destruction requires
a careful analysis of the bolometric magnitudes of the observed
stars, as discussed in Section 2. ii) Magnetic buoyancy might
offer a framework for interpreting both moderately Li-rich and
Li-poor stars. iii) Extensive Li processing is generally
accompanied by $^3$He destruction, which fact should be relevant
for reconciling the inventory of this nucleus with Big Bang and
stellar nucleosynthesis \citep{bs99}. iv) When the $^3$He
destruction is very effective, and thermohaline mixing might be
therefore inhibited \citep{den09}, magnetic buoyancy might still
offer a viable mechanism for explaining extra-mixing effects. v)
O-rich AGB stars might experience extra-mixing processes
differently than C-rich ones, with the tendency to favour higher
Li abundances. We remember that also AGB luminosities
\citep{guabus} and N abundances \citep{sl90} suggest differences
between MS-S and C(N) stars, according to which only rarely the
former class of objects will evolve to the latter one. More often,
they will preserve a different surface composition and a different
C/O ratio because they have different initial masses.

In a different paper of this volume \citep{palm2} we examine
the consequences of the mixing schemes here discussed for the isotopic
ratios of CNO elements and for $^{26}$Al.

\section*{Acknowledgments}
This research was supported by the Italian Ministry of Research
under contract PRIN2006-022731. The idea of mixing by magnetic
buoyancy was developed in common with G.J. Wasserburg and K.M.
Nollett.

%\end{multicols}


\begin{thebibliography}{}
% References are listed as in the following example, for more examples, please
% see the PASA Style Guide

\bibitem[Abia \& Isern(2000)]{abia}
Abia, C., Isern, J. 2000, ApJ, 536, 438

\bibitem[Andrews et al.(1988)]{and88}
Andrews, A.D., Rodon\'o, M., Lensky, J.L. et al. 1988, A\&A. 204, 177

\bibitem[Bergeat \& Chevallier(2005)]{bergeat}
Bergeat, J., Chevallier, L. 2005, A\&A, 429, 235

\bibitem[Boffin et al.(1993)]{boffin}
Boffin, H.~M.~J., Abia, C., Isern, J., Rebolo, R. 1993, A\&AS,
102, 361

\bibitem[Brown et al.(1989)]{brown}
Brown, J.~A., Sneden, C., Lambert, D.~L., Dutchover, E.~Jr. 1989,
ApJS, 71, 293

\bibitem[Busso et al.(2007a)]{busso07a}
Busso, M, Wasserburg, G.~J., Nollett, K.~M., Calandra, A. 2007a,
ApJ, 671, 802

\bibitem[Busso et al.(2007b)]{busso07b}
Busso, M., Guandalini, R., Persi, P., Corcione, L.,
Ferrari-Toniolo, M. 2007b, AJ, 133, 2310

\bibitem[Cameron \& Fowler(1971)]{camfow}
Cameron, A.~G.~W.; Fowler, W.~A. 1971, ApJ, 164, 111

\bibitem[Castilho(2000)]{cast}
Castilho B.~V. 2000, IAUS, 198, 331

\bibitem[Charbonnel(1994)]{charb94}
Charbonnel, C. 1994, A\&A, 282, 811

\bibitem[Charbonnel \& Do Nasci\-miento(1998)]{charb98}
Charbonnel, C., Do Nascimento, J.~D.~Jr. 1998, A\&A, 336, 915

\bibitem[Charbonnel \& Balachandran(2000)]{charb00}
Charbonnel, C., Balachandran, S.~C. 2000, A\&A, 359, 563

\bibitem[Denissenkov \& VandenBerg(2003)]{denvan}
Denissenkov, P.~A., VandenBerg, D.~A. 2003, ApJ, 593, 509

\bibitem[Denissenkov et al.(2009)]{den09}
Denissenkov, P.A., Pinsonneault, M., \& Mac Gregor K.B. 2009, ApJ
in press.

\bibitem[Denn et al.(1991)]{denn}
Denn, G.~R., Luck, R.~E., Lambert, D.~L. 1991, ApJ, 377, 657

\bibitem[Eggleon, Dearborn \& Lattanzio(2006)]{eggleton}
Eggleton, P.~P., Dearborn, D.~S.~P., Lattanzio, J.~C. 2006,
Science, 314, 1580

\bibitem[Gilroy(1989)]{gilroy}
Gilroy K.~K. 1989, ApJ, 347, 835

\bibitem[Gilroy \& Brown(1991)]{gilroybrown}
Gilroy K.~K., Brown J.~A. 1991, ApJ, 371, 578

\bibitem[Gratton et al.(2000)]{gratton}
Gratton, R.~G., Carretta, E., Matteucci, F., Sneden, C. 2000,
A\&A, 358, 671

\bibitem[Guandalini et al.(2006)]{gua06}
Guandalini, R., Busso, M., Ciprini, S., Silvestro, G., Persi, P.
2006, A\&A, 445, 1069

\bibitem[Guandalini \& Busso(2008)]{guabus}
Guandalini, R., Busso, M. 2008, A\&A, 488, 675

\bibitem[Guandalini, Tosti \& Busso(2008)]{irait}
Guandalini, R., Tosti, G., Busso, M. 2008, EAS Publications
Series, 33, 243

\bibitem[Kraft(1994)]{kraft}
Kraft R.~P. 1994, PASP, 106, 553

\bibitem[Lambert et al.(1980)]{lambert}
Lambert, D.~L., Dominy, J.~F., Sivertsen, S. 1980, ApJ, 235, 114

\bibitem[Luck \& Lambert(1982)]{luck}
Luck, R.~E., Lambert, D.~L. 1982, ApJ, 256, 189

\bibitem[Mallik(1999)]{mallik}
Mallik S.~V. 1999, A\&A, 352, 495

\bibitem[Melo et al.(2005)]{melo}
Melo, C.~H.~F., de Laverny, P., Santos, N.~C., Israelian, G.,
Randich, S., Do Nascimento, J.~D..~Jr., de Medeiros, J.~R. 2005,
A\&A, 439, 227

\bibitem[Nollett et al.(2003)]{nollett}
Nollett, K.~M., Busso, M., Wasserburg, G.~J. 2003, ApJ, 582, 1036

\bibitem[Nordhaus et al.(2008)]{nordhaus}
Nordhaus, J., Busso, M., Wasserburg, G.~J., Blackman, E.~G.,
Palmerini, S. 2008, ApJ, 684, L29

\bibitem[Palacios et al.(2003)]{palacios03}
Palacios, A., Talon, S., Charbonnel, C., Forestini, M. 2003, A\&A,
399, 603

\bibitem[Palacios et al.(2006)]{palacios06}
Palacios, A., Charbonnel, C., Talon, S., Siess, L. 2006, A\&A,
453, 261

\bibitem[Palmerini \& Busso(2008)]{palm1}
Palmerini, S., Busso, M. 2008, NewAR, 52, 412

\bibitem[Palmerini et al.(2008)]{palm2}
Palmerini, S., Busso, M., Guandalini R. 2008, PASA, this volume

\bibitem[Sackmann \& Boothroyd(1999)]{bs99}
Sackmann, I.-J., \& Boothroyd, A.I. 1999, ApJ 510, 217

\bibitem[Smith \& Lambert(1990)]{sl90}
Smith, V.V. \& Lambert, D.L. 1990, ApJS 72, 387

\bibitem[Sweigart \& Mengel(1979)]{sweigart}
Sweigart, A.~V., Mengel, J.~G. 1979, ApJ, 229, 624

\bibitem[Takeda et al.(2005)]{takeda}
Takeda, Y., Sato, B., Kambe, E., Izumiura, H., Masuda, Seiji,
Ando, H. 2005, PASJ, 57, 109

\bibitem[Uttenthaler et al.(2007)]{utt}
Uttenthaler, S., Lebzelter, T., Palmerini, S., Busso, M., Aringer,
B., Lederer, M.~T. 2007, A\&A, 471, L41

\bibitem[van Leeuwen(2007)]{hipparcos}
van Leeuwen, F. 2007, Hipparcos, the New Reduction of the Raw
Data. Astrophysics and Space Science Library, Vol. 350. Ed:
Dordrecht, Springer

\bibitem[Vanture et al.(2007)]{vanture}
Vanture, A.~D., Smith, V.~V., Lutz, J., Wallerstein, G., Lambert,
D., Gonzalez, G. 2007, PASP, 119, 147

\bibitem[Wasserburg \& Busso(2008)]{wasserburg}
Wasserburg, G.~J., Busso, M. 2008, AIPC, 1001, 295

\end{thebibliography}
\end{document}